\documentclass[nofootinbib,aps,11pt]{revtex4}
\usepackage[dvips]{graphicx}
\usepackage{amsmath}
\usepackage{bm}
\usepackage{times}
\usepackage{epsfig}
\usepackage{color}
\usepackage{amssymb}
\usepackage{textcomp}
\usepackage{url}

\begin{document}
\preprint{}
\vfill
\preprint{}
\vspace{7.0cm}
\title{\Large Grand Unification and Light Color-Octet Scalars at the LHC}
\vspace{4.0cm}
\author{Pavel Fileviez P{\'e}rez}
\email{fileviez@physics.wisc.edu}
\author{Ryan Gavin}
\email{rgavin@wisc.edu}
\author{Thomas McElmurry}
\email{mcelmurry@hep.wisc.edu}
\author{Frank Petriello}
\email{frankjp@physics.wisc.edu}
\affiliation{
{\small University of Wisconsin--Madison, Department of Physics} \\
{\small 1150 University Avenue, Madison, WI 53706, USA}}
\date{September 11, 2008}

\begin{abstract}
We study the properties and production mechanisms of color-octet
scalars at the LHC. We focus on the single production of both charged
and neutral members of an $({\bf 8,2})_{1/2}$ doublet through bottom
quark initial states. These channels provide a window to
the underlying Yukawa structure of the scalar sector.
Color-octet scalars naturally appear in grand unified
theories based on the $SU(5)$ gauge symmetry.  
In the context of adjoint $SU(5)$ these fields are expected to be light to satisfy constraints coming from unification and proton decay, and may have TeV-scale masses.  
One combination of their couplings is defined by the relation between the down-quark and charged-lepton Yukawa couplings.  Observation of these states at the LHC gives an {\it upper} bound on the proton lifetime if they truly arise from this grand unified theory.  We demonstrate that TeV-mass scalars can be observed over background at the LHC
using boosted top quark final states, and study how well the scalar
Yukawa parameters can be measured.
\end{abstract}
\maketitle

\section{Introduction}
Many motivations exist for physics beyond the Standard Model (SM)
of particle interactions.
Much of the research in high-energy physics for the past decades has been driven by the principle of naturalness.
This idea states that widely disparate scales in Nature, such as the
electroweak and Planck scales, should not exist without some symmetry
principle or dynamical mechanism to explain their ratio. Additional
motivations to believe in physics beyond the SM have recently been
found. These include the experimental discovery of neutrino masses
and the strong indication of TeV-scale dark matter. 

Recent study of physics beyond the SM has focused on the TeV scale,
both because of dark matter and because of the imminent start of the
experimental program at the Large Hadron Collider (LHC). Significant
effort has focused on brainstorming all possible scenarios that might
be observable at the LHC. Two possible questions to ask are the
following. Are there forms of new physics whose observability
at the LHC has not been studied?  Do current experimental data motivate
certain types of TeV-scale physics?  Several interesting answers to the
second question have recently been given. An example relevant to the
discussion here is the experimental absence of flavor-changing neutral
currents, which suggests that new physics should satisfy the principle
of minimal flavor violation (MFV). This states that the flavor structure
of physics beyond the SM should be completely determined by the SM Yukawa
structure~\cite{D'Ambrosio:2002ex}. Another question that arises is whether
measurements at the LHC can be used to understand details of high-scale
theories.  Both the seesaw mechanism for neutrino masses and the near
unification of SM gauge couplings suggest new physics near $10^{15}$ GeV.
Effort has been devoted to determine whether LHC results can indirectly
probe these high energies.

An interesting scenario of physics beyond the SM suggested by the 
above questions is the possibility of a scalar sector containing more than
the single Higgs doublet of the SM.  One motivation for considering extended
scalar sectors is purely phenomenological. Additional scalars can cause
large deviations in the branching fractions of the standard Higgs boson,
modifying experimental search strategies~\cite{higgsmod}. Their LHC signatures have not been
extensively studied. Another motivation is that certain seldom studied scalar
representations naturally satisfy the principle of minimal flavor violation (MFV).
A recent analysis showed that scalars transforming as either $(\mathbf1,\mathbf2)_{1/2}$ or $(\mathbf8,\mathbf2)_{1/2}$ under the gauge group $SU(3)_C \otimes SU(2)_L \otimes U(1)_Y$ can naturally have fermionic couplings proportional to the SM Yukawa structure~\cite{Manohar:2006ga}.
While the first representation is just that of the Higgs doublet,
the second denotes a color-octet scalar boson not found in the SM.

Color-octet scalars also potentially provide a window to high-scale physics. 
The unification of the gauge interactions is one of the main
motivations for physics beyond the SM, and it occurs naturally within 
Grand Unified Theories (GUTs).
Recently, in the context of adjoint $SU(5)$~\cite{adjoint}, the possibility of light color octets in agreement with the constraints coming from the unification of gauge couplings and proton decay was pointed out~\cite{adjoint1}. 
The couplings of these states may therefore probe physics at the GUT scale.  While observation of color-octet states 
at the LHC would not prove the existence of a GUT in Nature, it would offer the interesting possibility of correlating measurements 
from the LHC with future searches for proton decay.

Several studies regarding the discovery of color-octet scalars at the LHC have been performed~\cite{Manohar:2006ga,Gerbush:2007fe,Pheno-Octets,Octet-Higgs,Octet-Higgs2}.
The most promising channel appears to be pair production of either two neutral scalars
or a charged pair through the partonic channel $gg \to SS$, where $S$ denotes either
a neutral or charged state~\cite{Gerbush:2007fe}. Discovery of color-octet scalars of at least 1 TeV and
likely much higher is possible at the LHC. As the QCD couplings of the scalars to the
gluon are fixed by their octet representation, this mode has the advantage of being
model-independent. While this is a nice feature for discovery of the state,
it does not allow the structure of the Yukawa couplings to be measured. 
The situation is similar to the study of the top quark at the Tevatron.
The discovery mode $p\bar{p} \to t\bar{t}$ doesn't probe the underlying Yukawa structure
of the top quark. The measurement of $V_{t b}$ requires the study of single top production.

In analogy with single top production, we investigate the use of single production
of charged scalars through the partonic channel $bg \to S^{\pm}t$ to probe the
Yukawa structure of the $({\bf 8,2})_{1/2}$  scalar and the connection
to the relation between $Y_b$ and $Y_\tau$. It is likely that the dominant decay of
the scalar is through $S^{+} \to t\bar{b}$, leading to the production process
$pp \to S^{+} \bar{t} + S^- t\to t\bar{t}\bar{b}+t\bar{t}b$. We expect the dominant background to
this signal to be $t\bar{t}j$, as we discuss later in more detail. We perform an
analysis indicating that the signal can be seen over background at the LHC, and
identify the relevant kinematic features that make this possible. We also study
neutral scalar resonance production through the partonic channel $b\bar{b} \to S \to t\bar{t}$,
which can become sizable for a large scalar Yukawa coupling to the bottom quark.
This channel provides another window into the Yukawa sector of the color-octet scalar.
We illustrate that this can also be observed over background at the LHC.  We discuss how well the 
scalar Yukawa couplings can be measured in these channels. 

Our paper is organized as follows. We present our notation and setup, and explain
how color-octet scalars provide a window to GUT-scale physics, in Section~\ref{sec:setup}.  We discuss the various production and decay mechanisms 
for color-octet scalars in Section~\ref{sec:prod}.  In Section~\ref{sec:sim} we describe the details of our simulation procedure.
We apply this analysis procedure to the study of single charged scalar production
and neutral scalar production through the $b\bar{b}$ initial state in Section~\ref{sec:analysis},
and demonstrate how to separate signal from background in these two channels.
We conclude in Section~\ref{sec:conc}.  Analytic expressions for the production and decay 
modes used in the analysis are given in the Appendix.
\section{MFV, grand unification, and light color octets \label{sec:setup}}
We first present some basic results relevant for studying the scalar octets
at the LHC. We only discuss issues which directly affect our analysis; details
on other aspects of color octets can be found in Ref.~\cite{Manohar:2006ga}.
\subsection{Minimal flavor violation and light octets}
We consider an extension of the Standard Model where the scalar sector is composed of the SM Higgs, $H\sim(\mathbf1,\mathbf2)_{1/2}$, and a color octet, $S\sim(\mathbf8,\mathbf2)_{1/2}$.
In this case the extra Yukawa interactions due to the presence
of the octet are given by
\begin{equation}
{\cal L}_Y = \bar{d}_R \ \Gamma_D \ S^\dagger \ Q_L \ + \ \bar{u}_R \ \Gamma_U \ Q_L^\alpha \ S^\beta \ \epsilon_{\alpha \beta} \ + \ \text{h.c.},
\label{int1}
\end{equation}
where
\begin{equation}
S =
\left(
\begin{array} {c}
 S^+  \\
 S^0
\end{array} \right)
=
\left(
\begin{array} {c}
 S^+  \\
 \frac{S_R^0 \ + \ i \ S_I^0}{\sqrt{2}}
\end{array} \right)
=S^a T^a,
\end{equation}
$a=1,\ldots,8$ and $T^a$ are the $SU(3)$ generators.
In the physical basis,
\begin{equation}
\mathcal L_Y=\begin{aligned}[t]
&\bar d\left[P_L\left(D_R^\dagger\Gamma_DU_L\right)-P_R\left(D_L^\dagger\Gamma_U^\dagger U_R\right)\right]S^-u+\bar u\left[P_R\left(U_L^\dagger\Gamma_D^\dagger D_R\right)-P_L\left(U_R^\dagger\Gamma_U D_L\right)\right]S^+d\\
&+\frac{S_R^0}{\sqrt2}\bar d\left[P_L\left(D_R^\dagger\Gamma_D D_L\right)+P_R\left(D_L^\dagger\Gamma_D^\dagger D_R\right)\right]d+\frac{S_R^0}{\sqrt2}\bar u\left[P_L\left(U_R^\dagger\Gamma_U U_L\right)+P_R\left(U_L^\dagger\Gamma_U^\dagger U_R\right)\right]u\\
&+i\frac{S_I^0}{\sqrt2}\bar d\left[-P_L\left(D_R^\dagger\Gamma_DD_L\right)+P_R\left(D_L^\dagger\Gamma_D^\dagger D_R\right)\right]d+i\frac{S_I^0}{\sqrt2}\bar u\left[P_L\left(U_R^\dagger\Gamma_U U_L\right)-P_R\left(U_L^\dagger\Gamma_U^\dagger U_R\right)\right]u,
\end{aligned}
\end{equation}
where the matrices $U_L$, $U_R$, $D_L$ and $D_R$ diagonalize the mass matrices for quarks.  $S^{\pm}$ denotes the charged octet scalar and $S^0_{R,I}$ 
are respectively the $CP$-even and $CP$-odd neutral scalars.
If we assume minimal flavor violation~\cite{Manohar:2006ga}, then
\begin{equation}
\Gamma_U = \eta_U Y_U \;\text{and} \;\Gamma_D = \eta_D Y_D.
\label{etadef}
\end{equation}
In this case the physical interactions are
\begin{equation}\label{lagrangian}
\mathcal L_Y^\text{MFV}=\begin{aligned}[t]
&\frac{\sqrt2}v\bar d\left(P_L\eta_Dm_DV_\text{CKM}^\dagger-P_R\eta_UV_\text{CKM}^\dagger m_U\right)S^-u\\
&+\frac{\sqrt2}v\bar u\left(P_R\eta_DV_\text{CKM}m_D-P_L\eta_Um_UV_\text{CKM}\right)S^+d\\
&+\eta_D\frac{m_D}vS_R^0\bar dd+\eta_U\frac{m_U}vS_R^0\bar uu+i\eta_D\frac{m_D}vS_I^0\bar d\gamma_5d-i\eta_U\frac{m_U}vS_I^0\bar u\gamma_5u,
\end{aligned}
\end{equation}
where $V_\text{CKM}$ is the Cabibbo-Kobayashi-Maskawa matrix, $v$ is the SM Higgs vev, $u$ and $d$ are respectively the SM up- and down-type quarks, and $m_U,m_D$ are their masses.
$\eta_U$
and $\eta_D$ are parameters that describe the strength of the scalar couplings to matter.

The mass splittings between the charged state and the neutral members
of the $({\bf 8,2})_{1/2}$ depend on the details of the potential describing
their self-interactions and their coupling to the regular Higgs doublet.
If the splitting is sufficiently large, decays such as $S^{\pm} \to S^0 W^{\pm}$
are allowed, where $S^0$ denotes one of the neutral scalars. Otherwise, the charged
scalar will decay predominantly via $S^{+} \to t\bar{b}$ due to the
couplings in Eq.~\ref{lagrangian}. An analysis of the scalar-potential parameter
space allowed by current experimental constraints reveals that scalar cascade decays
are unlikely~\cite{Gerbush:2007fe}. We will assume in our study that the
charged scalar decays only to $tb$ pairs.
\subsection{Adjoint $SU(5)$ and light color octets}
The existence of the color octets mentioned above can also be understood
in the context of the simplest grand unified theories based
on $SU(5)$~\cite{GG}.
For a review on grand unified theories and their phenomenological aspects, see Ref.~\cite{review}.
In Ref.~\cite{GJ} it was realized that in order to have a consistent relation between the charged lepton and down quark masses in the context of renormalizable $SU(5)$ theories, a new Higgs in the \textbf{45} representation must be introduced.
In these models the Higgs sector
is composed of the representations $\bm{5_H}$, $\bm{24_H}$ and $\bm{45_H}$, and the relevant Yukawa
interactions are given by
\begin{eqnarray}
- S_{\rm Yukawa} &=&
\int d^4 x
\Bigg( Y_1 \ 10 ~\bar{5} ~5^*_H
\ + \
Y_2 10 ~\bar{5} ~45^*_H
\ + \ Y_3 \ 10~ 10 ~ 5_H  \
+ \
Y_4 \ 10 ~10 ~45_H \Bigg) \  + \ \text{h.c.},
\label{Yukawa}
\end{eqnarray}
where $\bar{5}=(d^C, l)_L$, and $10=(u^C, Q, e^C)_L$.  The masses of the SM charged fermions are
\begin{eqnarray}
M_D & = & Y_1 \frac{v_5^*}{\sqrt{2}} \ + \ 2 \ Y_2 \frac{v_{45}^*}{\sqrt{2}}~, \\
M_E & = & Y_1^T \frac{v_5^*}{\sqrt{2}} \ - \ 6 \ Y_2^T \frac{v_{45}^*}{\sqrt{2}}~, \\
\text{and} \;\;M_U & = & 4 \left( Y_3 + Y_3^T \right) \frac{v_5}{\sqrt{2}} \ - \ 8 \ \left( Y_4 - Y_4^T \right)
\frac{v_{45}}{\sqrt{2}}\equiv M_{5}^U \ + \ M_{45}^U,
\label{gutmass}
\end{eqnarray}
with $v_5/\sqrt{2}=\langle {5_H} \rangle$, and $v_{45}/\sqrt{2} = \langle {45_H} \rangle^{15}_1 = \langle {45_H} \rangle^{25}_2 = \langle {45_H} \rangle^{35}_3$.

The color octet studied in Ref.~\cite{Manohar:2006ga} lives in the \textbf{45} representation of the $SU(5)$ theory, $S\subset 45_H$.  We note that $S=\Phi_1$ in the notation of 
Ref.~\cite{adjoint,adjoint1}.
Using the above Yukawa interactions we find the following interactions between the SM fermion and the octet:
\begin{eqnarray}
\bar{d}_R \ \frac{(M_E - M_D^T)}{2 \sqrt{2} v_{45}^*} \ S^\dagger \ Q_L, \ \qquad
\ \qquad \bar{u}_R \ \frac{\sqrt{2} M_{45}^U}{v_{45}}  \ Q_L^\alpha \ S^\beta \epsilon_{\alpha \beta},
\end{eqnarray}
where
\begin{equation}
S^{i \alpha}_j= (45_H)^{i \alpha}_j - \frac{1}{3} \ \delta^i_j \ (45_H)^{m \alpha}_m,
\end{equation}
the indices $i,j=1,2,3$ and $\alpha=4,5$.  The couplings in Eq.~(\ref{int1}) are then defined by
\begin{equation}
\Gamma_u = \sqrt{2} \frac{M_{45}^U}{v_{45}}, \qquad \text{and} \qquad \Gamma_d = \frac{M_E - M_D^T}{2 \sqrt{2} v_{45}^*}.
\label{guteta}
\end{equation}
As pointed out in Ref.~\cite{Steve}, this model has only one light Higgs doublet, and $v_5^2 \ + \ v_{45}^2 = v^2$.

Recently, a grand unified theory has been proposed where the neutrino masses
are generated through the Type I and Type III seesaw mechanisms~\cite{adjoint}.  This predicts
the existence of light color-octet scalars. This issue has been studied in detail
in Ref.~\cite{adjoint1}, where it was concluded that in order to satisfy
the constraints coming from proton decay and unification of gauge interactions,
the color octets must be very light.
The upper bound on the mass of this color octet is $m_S< 4.4 \times 10^5$~GeV~\cite{adjoint1}.
This result indicates it may be possible to produce these exotic fields
at the LHC.

In order to investigate the couplings of the octets at the LHC and understand their connection
to the GUT theory, we assume that at low energy the MFV hypothesis
is valid.  Neglecting mixings, we find
\begin{equation}
\eta_D = \frac{v}{4 v_{45}^* m_{b}} \left( m_{\tau} - m_{b} \right).
\end{equation}
Notice that since $m_b(M_Z)=2.89$ GeV, $m_\tau (M_Z)=1.746$ GeV~\cite{German}, and $v_{45} < v$, the coupling $\eta_d$ can be significantly larger than unity.   Measurement of 
$\eta_D$ fixes $v_{45}$, the vev of $\bm{45_H}$.  Determination of $\eta_U$ then probes the Yukawa coupling $Y_4$ that parametrizes the interaction between $\bm{45_H}$ and 
up-type quarks, as can be seen using Eqs.~\ref{gutmass} and~\ref{guteta}.

\begin{figure}[htbp]
   \centering
   \includegraphics[width=0.6\textwidth,angle=0]{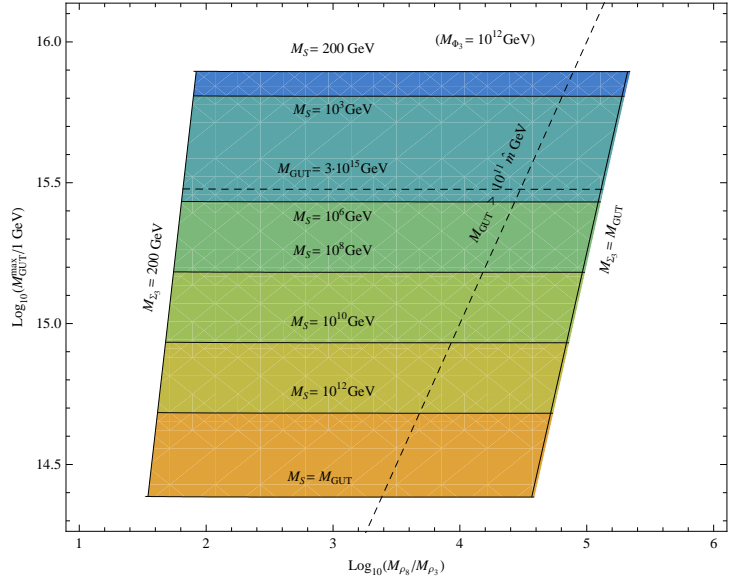}
   \caption{\label{gutplot} Relationship between the color-octet scalar mass $m_S$, the maximum GUT scale $M^{\rm max}_{{\rm GUT}}$, and various other mass scales of the $SU(5)$ theory.  The horizontal dashed line on this plot indicates the current constraint from non-observation of proton decay, while the diagonal dashed line denotes the requirement on GUT parameters 
needed to obtain successful leptogenesis.  Regions in the upper-left portion of the figure are allowed by these constraints.  Contours of constant color-octet scalar mass are horizontal lines in this figure. For further details see Refs.~\cite{adjoint,adjoint1,Steve}.}
\end{figure}

In order to ascertain that the color-octet scalar is coming from a GUT, information from other experiments is required.  
Observation of proton decay would constitute strong evidence of grand unification in 
Nature.  Remarkably, an {\it upper} bound on the proton lifetime is predicted by measurement of the color-octet mass at the LHC.  This is illustrated in Fig.~\ref{gutplot}, which shows the relationship 
between the color-octet scalar mass $m_S$, the GUT scale $M_{{\rm GUT}}$, and various other mass scales of the $SU(5)$ theory (for more details see Refs.~\cite{adjoint,adjoint1}).  The horizontal dashed line on this plot indicates the current constraint from non-observation of proton decay, while the diagonal dashed line denotes the requirement on GUT parameters 
needed to obtain successful leptogenesis~\cite{Steve}.  Regions in the upper-left portion of the figure are allowed by these constraints.   Contours of constant color-octet scalar mass are horizontal lines in this figure.  The important point to note is that the color-octet scalar mass fixes the maximum possible GUT scale, $M^{\rm max}_{{\rm GUT}}$.  This determines an upper bound on the 
proton lifetime, as the dominant contributions to this process are the super-heavy $X$ and $Y$ bosons of the $SU(5)$ theory, whose 
couplings are fixed by unification.  For example, if $m_S=1\;{\rm TeV}$, $M^{\rm max}_{{\rm GUT}}\approx 6 \times 10^{15}\;{\rm GeV}$, leading to the following bounds on the partial lifetimes:
\begin{eqnarray}
\tau(p\to K^+ \bar{\nu}) &\leq& 4 \times 10^{36}\; {\rm years}; \nonumber \\
\tau(p\to \pi^+ \bar{\nu}) &\leq& 1 \times 10^{35}\; {\rm years}; \nonumber \\
\tau(p\to e^+ \pi^0) &\leq& 5 \times 10^{34}\; {\rm years}.
\end{eqnarray}
Measurement of $m_S$ at the LHC then gives an unavoidable prediction for the rate of proton decay.

\section{Octet decays and production mechanisms at the LHC \label{sec:prod}}
Several partonic production mechanisms can lead to color-octet scalar production at the LHC.  Pair production processes such as $gg \to S^+S^-,S_R S_R,S_I S_I$, which
depend only on the QCD coupling, were studied in Ref.~\cite{Gerbush:2007fe}.  The gluon-initiated single-production processes $gg\to S_{R,I}$, studied in Ref.~\cite{Gresham:2007ri}, can
be important for large $\eta_U$.  If $\eta_D$ is larger than unity, then $b\bar{b}\to S_{R,I}$ becomes significant.  Charged scalars can be singly produced via the partonic
interaction $g\bar{b} \to t S^{-}$ and its charge conjugate.
Pair production of a charged and a neutral scalar is possible via an intermediate $W$ boson, for example through the partonic process $u\bar{d} \to W^+ \to S^{+}S_{R,I}$, and is fixed by the $SU(2)_L$ quantum number of the scalar doublet.
We focus here on the single-production processes $b\bar{b}\to S_{R,I}$ and $g\bar{b} \to t S^{-}$ which allow
measurements of $\eta_U$ and $\eta_D$.

As mentioned previously, we focus on top-quark decays of the scalars, $S^+\to t\bar{b}$ and $S_{R,I} \to t\bar{t}$, as a study has indicated that the mass splitting between the scalar states
is likely not large enough for cascade decays such as $S^{\pm} \to S_{R,I}W^{\pm}$ to be important~\cite{Gerbush:2007fe}.  If both $\eta_U$ and $\eta_D$ are small, then
the scalars might be stable on collider length scales, leading to missing energy and highly ionizing tracks.  Scalar bound states can also occur in this case near pair-production thresholds.  We do
not further pursue these possibilities here.

The cross sections for the processes $pp \to S^+S^-$ and $pp \to tS^{-}+\bar{t}S^+$ are simple to derive given the interactions in Eq.~(\ref{lagrangian}).  We
present them below in Fig.~\ref{csecs} as a function of the scalar mass $m_S$ for $\eta_U=\eta_D=1$.  These are leading-order results obtained using the CTEQ6L1 parton distribution
functions~\cite{Pumplin:2002vw}.  Two things should be noted from these results.
First, the cross section for $m_S \sim 1\, {\rm TeV}$ is sufficiently large to allow study of the single scalar production mode.
Second, the $S^+S^-$ production process decreases more quickly with rising scalar mass.
This is simply due to the phase space behavior of each process.  The di-scalar production process falls off as $\hat{s}-4m_S^2$, while
the single production decreases as $\hat{s}-(m_S+m_t)^2$, where $\hat{s}$ is the partonic center-of-mass energy.  The single charged scalar production mode
may therefore extend the search reach at high $m_S$, although we do not pursue this possibility here.

We also study the resonant production of $S_{R,I}$ through the partonic process $b\bar{b} \to S_{R,I} \to t\bar{t}$, which can become important for large values
of $\eta_D$.  For illustration we plot this cross section also in Fig.~\ref{csecs} assuming $\eta_D=40$ and $\eta_U=1$.  Analytic expressions for the $bg \to S^{\pm} t$,
$b\bar{b} \to S_{R,I}$, and $gg \to S^+S^-$ partonic cross sections, as well as for the scalar decays, are given in the Appendix.

\begin{figure}[htbp]
   \centering
   \includegraphics[width=0.4\textwidth,angle=90]{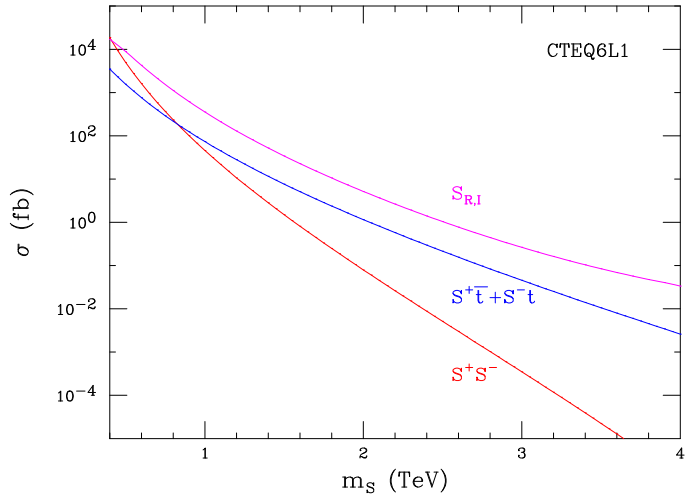}
   \caption{\label{csecs} Inclusive cross sections for the processes $pp \to S^+S^-$, $pp \to tS^{-}+\bar{t}S^+$, and $pp\to S_{R,I}$ through bottom quark fusion at the LHC.
We have set the renormalization and factorization scales to $\mu=m_t+m_S$ for single scalar production, to $\mu=2m_S$ for di-scalar production, and to $\mu=m_S$ for single scalar production, and have used the CTEQ6L1 parton distribution functions.}
\end{figure}

\section{Details of simulation \label{sec:sim}}
To discuss production and measurement of color-octet scalars at the LHC, we must determine how well the signal can be
observed over background.  We describe here our procedure for estimating this.

We implement the charged and neutral scalars in MadEvent~\cite{Alwall:2007st} using Feynman rules derived from Eq.~(\ref{lagrangian}).  We set the renormalization and factorization scales on an event-by-event basis using $\mu_{R}^2=\mu_{F}^2=m_{t}^2+\sum p_{T}^2$, where the sum runs over all objects in the final state including missing energy.  After the scalar decay, the final state is $t\bar{t}b+t\bar{t}\bar{b}$.  We decay the top quarks fully using the \verb|DECAY| routine of MadEvent, which does not preserve spin correlations.  MadEvent allows spin correlations to be kept throughout the entire decay chain; we have checked on a smaller sample than that used in our full analysis, that they have a negligible effect given the cuts we use.  We
set $m_S=1 \,{\rm TeV}$ to illustrate the analysis procedure.  We discuss later our simulation of the backgrounds.  We compute all cross sections at leading order in QCD; while
the next-to-leading order results for the backgrounds we will later consider are known, those for our signal processes are not.  This is a conservative choice, as this procedure will tend to underestimate the obtainable statistical significance $S/\sqrt{B}$ if the QCD corrections increase both cross sections roughly equally.

The above procedure leads to sets of parton-level events for both signal and background.  We now describe our procedure of applying a simple detector simulation to these events 
and finding a set of cuts to extract the signal.
\subsection{Detector simulation}
We perform a simple estimate of detector effects by
applying the following four actions to our parton-level events.

\begin{itemize}

\item We model the acceptances of LHC detectors by discarding objects with $|\eta| > 2.5$.

\item To simulate the effect of imperfect energy resolution, we subject each particle in the final state to a Gaussian smearing of either its energy $E$ or its transverse momentum $p_T$.  We use the following smearing parameters.
\begin{itemize}
\item jet-like objects (light quarks, $b$ quarks, gluons):  $\;\;\frac{\delta E}{E} = \frac{0.8}{\sqrt{E/{\rm GeV}}}\oplus 0.03$.
\item electrons:  $\;\;\frac{\delta E}{E} = \frac{0.1}{\sqrt{E/{\rm GeV}}}\oplus 0.007$.
\item muons:  $\;\;\frac{\delta p_{T}}{p_{T}} = 0.15 \: p_{T}/{\rm TeV}\oplus 0.005$.
\end{itemize}
These values are consistent with LHC expectations~\cite{tdrs}.  After all final state objects have been smeared, the transverse momentum is recalculated.  The $p_{x}$ and $p_{y}$ components of the missing energy are set to those values which ensure momentum conservation.

\item We combine partons using a cone algorithm with $\Delta R=0.4$.  These combined objects are then considered jets.

\item We mimic LHC capabilities for tagging heavy flavor jets by assuming $b$-jets are tagged with 60\% efficiency, and that other jets
fake $b$-jets with a 3\% fake rate.

\end{itemize}
An alternative to the above smearing of parton-level events is to run these events through the PYTHIA parton-shower~\cite{Sjostrand:2006za} and PGS detector simulation~\cite{pgs}.  We have tested that using this other technique of showering the parton-level events with PYTHIA and then using PGS to reconstruct observable objects leads to qualitatively similar results as found with the 
procedure described above; a slight degradation of our figure of merit $S/\sqrt{B}$ occurs with the alternate procedure.  We present results using the smearing technique described above due to its simplicity and transparency.

\subsection{Top reconstruction}\label{ss:recon}
Since the final state contains two top quarks, their reconstruction becomes
an important part of the analysis.  With $m_S=1\,{\rm TeV}$, both tops in the signal
sample will be highly boosted.  These should then decay into a collimated $b$ quark
and $W$ boson.  This will be instrumental in constructing a top-finding
algorithm.  We study the reconstruction only of events where one top quark decays
leptonically or both decay hadronically, and we do not attempt to reconstruct $\tau$ leptons.
If there is a single lepton present in the final state, the $W$ can be reconstructed using the lepton and missing energy and then combined with a $b$-jet to reconstruct the top quark. If there are no leptons in
the final state, one can look for jets or combinations of jets with an invariant mass
near that of the top.  Several recent studies have used similar methods for
reconstructing highly boosted top quarks~\cite{Agashe:2006hk,Barger:2006hm,Lillie:2007yh,Skiba:2007fw,Baur:2007ck,Thaler:2008ju,Kaplan:2008ie}.
We describe below the details of our semi-leptonic and hadronic top algorithms.
\subsubsection{Semi-leptonic top reconstruction}
The semi-leptonic algorithm is applied when there is only one lepton in the final state, either an electron or muon.  Motivated by the significant boost of the top quark,
we search for the jet with the smallest $\Delta R = \sqrt{\Delta \phi^2+\Delta\eta^2}$ separation from the lepton.  We do not require the jet to be $b$-tagged.  If the separation between the lepton and jet is
greater than $\Delta R_{{\rm max}}=0.6$, the event is passed on to the hadronic top reconstruction algorithm.  Otherwise, the lepton and missing energy are combined such
that the longitudinal component of the missing energy is constrained by the $W$ mass.  There is a two-fold ambiguity in determining the longitudinal momentum;
the degeneracy is broken by choosing the solution with the smallest $\Delta R$ separation between the reconstructed $W$ and the previously chosen jet.  If the
invariant mass of the combined jet, lepton, and missing energy falls within a window $171 \pm 50$ GeV, the combined object is labeled a top quark.
\subsubsection{Hadronic top reconstruction}
When no leptons are present after the semi-leptonic algorithm is performed, a purely hadronic algorithm attempts to reconstruct other top quarks in the event.
If the invariant mass of any single jet falls within the top mass window $171 \pm 50$ GeV, indicating the possibility that all of the top decay products have been combined by the jet algorithm, this jet is tagged as a top quark.
Next, the two jets with the smallest $\Delta R$ separation are found, again imposing a maximum value $\Delta R_{{\rm max}}=0.8$.
If the invariant mass of these two jets falls within the top mass window, they are combined and tagged as a top quark.
If the invariant mass of the two jets falls within the range 40--121 GeV, indicating that they may be the decay products of a $W$ boson, they are combined provisionally, and the jet with the smallest $\Delta R_{W,j}<\Delta R_{{\rm max}}$ is found.
If the invariant mass of the $W$ candidate and the third jet falls within the top mass window, the three jets are combined and tagged as a top quark; otherwise, the combination of the first two jets to form the $W$ candidate is undone.
This algorithm continues until no further top quarks are found.
\section{Analysis \label{sec:analysis}}
\subsection{Single charged scalar production}
We now present results obtained by applying the above analysis procedure to the analysis of single charged scalar production via the partonic process $gb \to S^{\pm}t \to ttb$, where $b,t$ can
represent either quarks or antiquarks.  We discuss the kinematic features that allow separation of signal from background, and estimate the obtainable statistical significance.  We use $S/\sqrt{B}$ as our statistical measure to determine whether the signal is observable over the statistical fluctuation in the expected background. 

The signal is proportional to the coupling $|\eta_U|^4/\Gamma_S\sim |\eta_U|^2$.
We set $\eta_D=\eta_U=1$ in this analysis.  Previous studies of semi-leptonic $t\bar{t}$ events at the LHC by the ATLAS and CMS collaborations have indicated
that the $t\bar{t}$ signal can be separated from the backgrounds of $WW$, $WZ$, $ZZ$, $W+{\rm jets}$, and $Z+{\rm jets}$, finding $S/B=65\,(27)$ with $10\,(1)\,{\rm fb}^{-1}$~\cite{tdrs,Barger:2006hm}.
We therefore focus only on the $t\bar{t}j$ background when reconstructing semi-leptonically decaying $t\bar{t}$ pairs, assuming that the same holds true with the addition of an extra jet.   Backgrounds to the fully hadronically decaying top pair are more dangerous, as QCD multi-jet production can lead to the same final state.  Studies of fully hadronic $t\bar{t}$ reconstruction by ATLAS and CMS have found that 
$S/B \approx 1/3$ is achievable with $1\,{\rm fb}^{-1}$~\cite{tdrs} .  Several recent studies have attempted to distinguish high invariant mass $t\bar{t}$ pairs from QCD backgrounds using jet mass and substructure differences~\cite{Thaler:2008ju,Kaplan:2008ie}, and have had some success.
For example, the study of Ref.~\cite{Kaplan:2008ie} found that $S/B \approx 1$ is achievable by decomposing a hadronically decaying top quark into sub-jets and using the reconstruction of two of these to $M_W$; this is similar to our technique described in Section~\ref{ss:recon}.  Simulation of these pure
QCD backgrounds is right now rather uncertain without LHC data to validate currently used Monte Carlo programs.  In our analysis we
simulate only the $t\bar{t}j$ background to the fully hadronic mode, and present two sets of results: one using only semi-leptonically reconstructed events, and another using both
semi-leptonic and hadronic events.  The actual LHC capability will likely be something intermediate between these two limits when QCD multi-jet backgrounds are included.  As will become clear in the following discussion,
the kinematic region of interest consists of three well-separated top and bottom jets, which validates our use of the $t\bar{t}j$ hard matrix elements without matching to a
parton shower.  We have checked this by passing our parton-level events through PYTHIA and PGS and finding similar results.

Imposing only the cuts $|\eta|<2.5$ and $p_T>20$ GeV
for $b$ and light jets, the signal cross section $pp \to S^+ \bar{t}+S^-t \to t\bar{t}b+t\bar{t}\bar{b}$ is 90 fb, while the $pp \to t\bar{t}+{\rm jet}$ background is 526 pb.  Signal and background events were generated using MadEvent, and then decayed using the \verb|DECAY| function.  The analysis was performed on $10^5$ signal events and $5 \times 10^6$ $t\overline{t}j$ events.  
Scales were set on an event-by-event basis such that $\mu_{R}^2=\mu_{F}^2=m_t^2+\sum p_{T}^2$, where the sum runs over all particles in the final state including missing energy.

\begin{figure}[htb]
\hspace{-0.1cm}
\includegraphics[angle=90,width=3.0in]{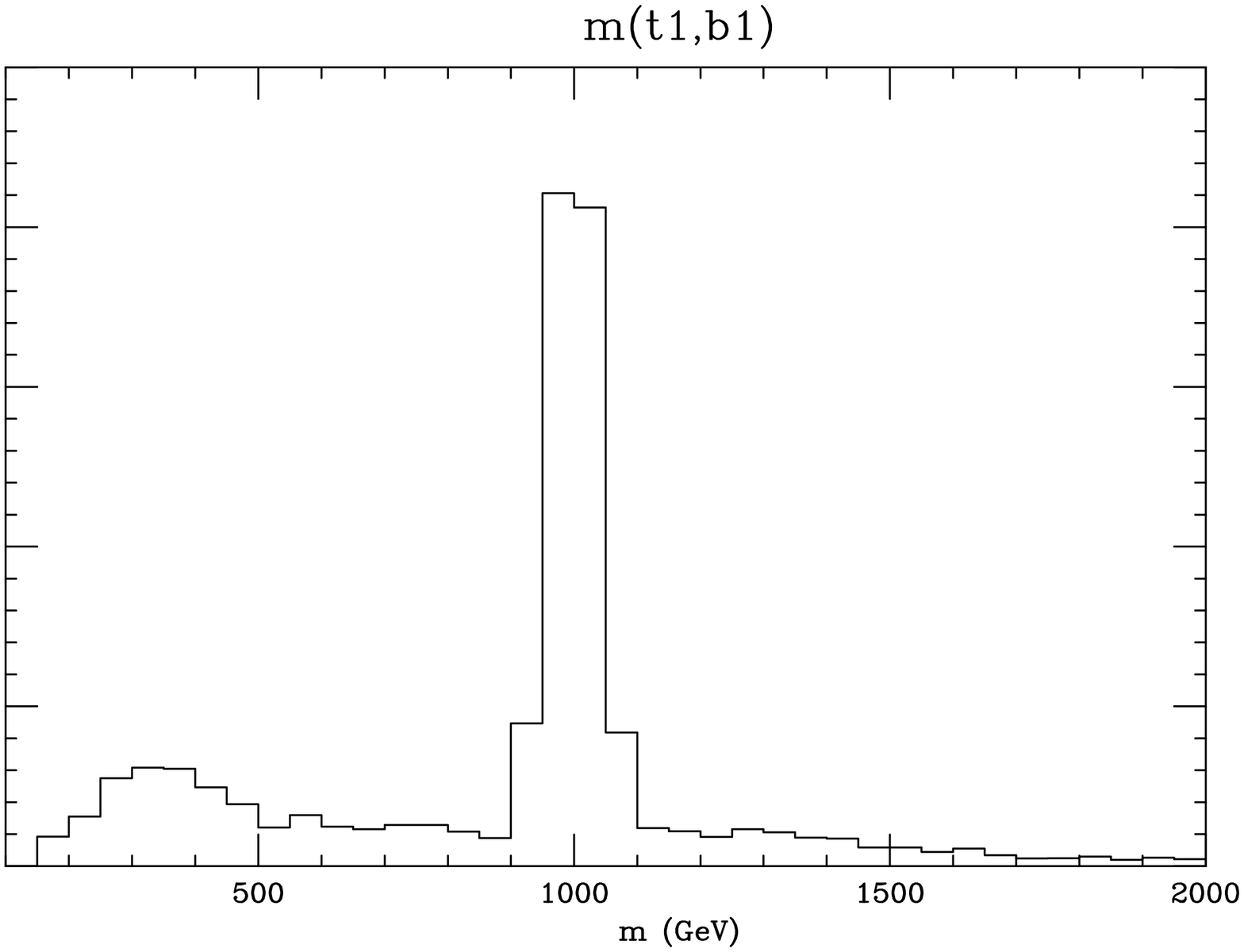}
\includegraphics[angle=90,width=3.0in]{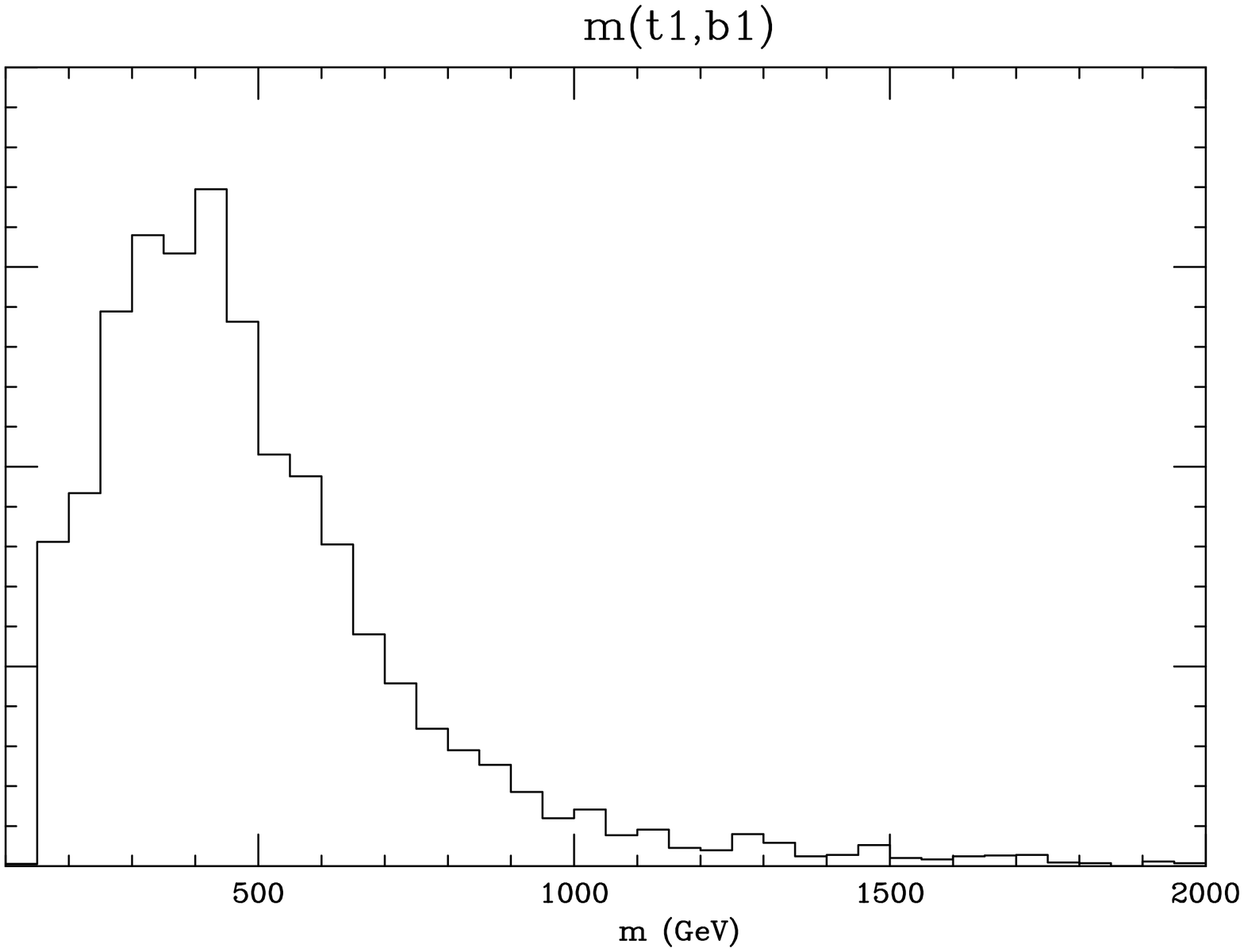}
\caption{ Invariant mass $M_{b_1,t_1}$ for the charged octet signal (left panel) and the $t\bar{t}j$ background (right panel).  Included are the two top tags, one $b$-tag, and
$H_T>1000\;{\rm GeV}$ cut of Table~\ref{kincuts}.
The normalization of the vertical axis is arbitrary.
All scales on each plot are linear.}
 \label{mplot}
\end{figure}

We first impose the following set of cuts motivated by the kinematics of the signal process.
We demand that our top reconstruction algorithm find two top quarks, and in addition require another $b$-tagged jet.
We expect the top quark recoiling against the scalar to be very energetic, and we also
expect the scalar decay products to be highly boosted.  We also expect the signal to have a large $\hat{s}$.  This motivates the series of kinematic cuts
presented in the upper half of Table~\ref{kincuts}.  We have defined $H_T$ as the scalar sum of the $E_{T}$ of all visible final state objects and the $p_T$ of the missing energy.  $t_1$ and
$b_1$ are respectively the hardest top-tagged and $b$-tagged jets, while $t_2$ is the softer of the two top tags.  We note that as this state can be discovered in the di-scalar
channel, we can tune our cuts to the value of $m_S$ measured in this other mode, as we have done here for $m_S=1\,{\rm TeV}$ in the invariant mass cut $900\:{\rm GeV}<M_{b_1,t_1}<1100\:{\rm GeV}$.  The size of this interval is larger than the expected calorimeter resolution~\cite{tdrs}.  These numbers are for $300 \,{\rm fb}^{-1}$ of integrated luminosity.   For illustration of the
kinematic difference between signal and background, the  $M_{b_1,t_1}$ distribution for both with arbitrary normalization is shown in Fig.~\ref{mplot}.

\begin{table}[htbp]
\centering
\begin{tabular}{c c c c c}
\hline\hline
Cut & $S_l$ & $S_h$ & $B_l$ & $B_h$\\ [1ex]
\hline
$H_T > 1000\:{\rm GeV}$ & $4940$ & $7820$ & $8.00\times 10^5$ & $1.26\times 10^6$ \\ [1ex]
$\exists \,t_1,t_2,b_1$ & $93$ & $283$ & $9610$ & $2.85\times 10^4$ \\[1ex]
$900\:{\rm GeV}<M_{b_1,t_1}<1100\:{\rm GeV}$ & $57$ & $158$ & $480$ & $835$ \\ [1ex]
\hline
$\Delta R_{b_1 t_1}, \Delta R_{b_1 t_2}< 3.0$ & 41 & 126 & 127 & 230\\ [1ex]
${\rm cos}\, \theta_{t_1 b_1}<0.5,{\rm cos}\, \theta_{t_2 b_1}<0.7$  & 34 & 111 & 54 & 144 \\ [1ex]
\hline
\end{tabular}
\caption{\label{kincuts} Effects of cuts on cross sections; the upper half includes cuts motivated by the energy released in the interaction, while the lower half is motivated by the event topology.  $S_l$ and $S_h$ denote the expected numbers of semi-leptonic and hadronic signal events after cuts for $300 \,{\rm fb}^{-1}$ of integrated luminosity; $B_l$ and $B_h$ denote the same for the background.}
\end{table}

\begin{figure}[htbp]
   \centering
   \includegraphics[width=0.25\textwidth,angle=0]{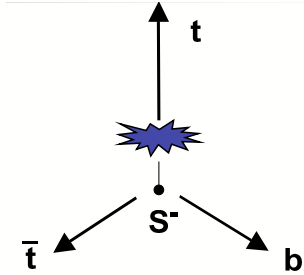}
   \caption{``Mercedes Benz" topology of the charged scalar signal process.}
   \label{MB}
   \vspace{-0.1cm}
\end{figure}

\begin{figure}[htb]
\hspace{-0.1cm}
\includegraphics[angle=90,width=3.0in]{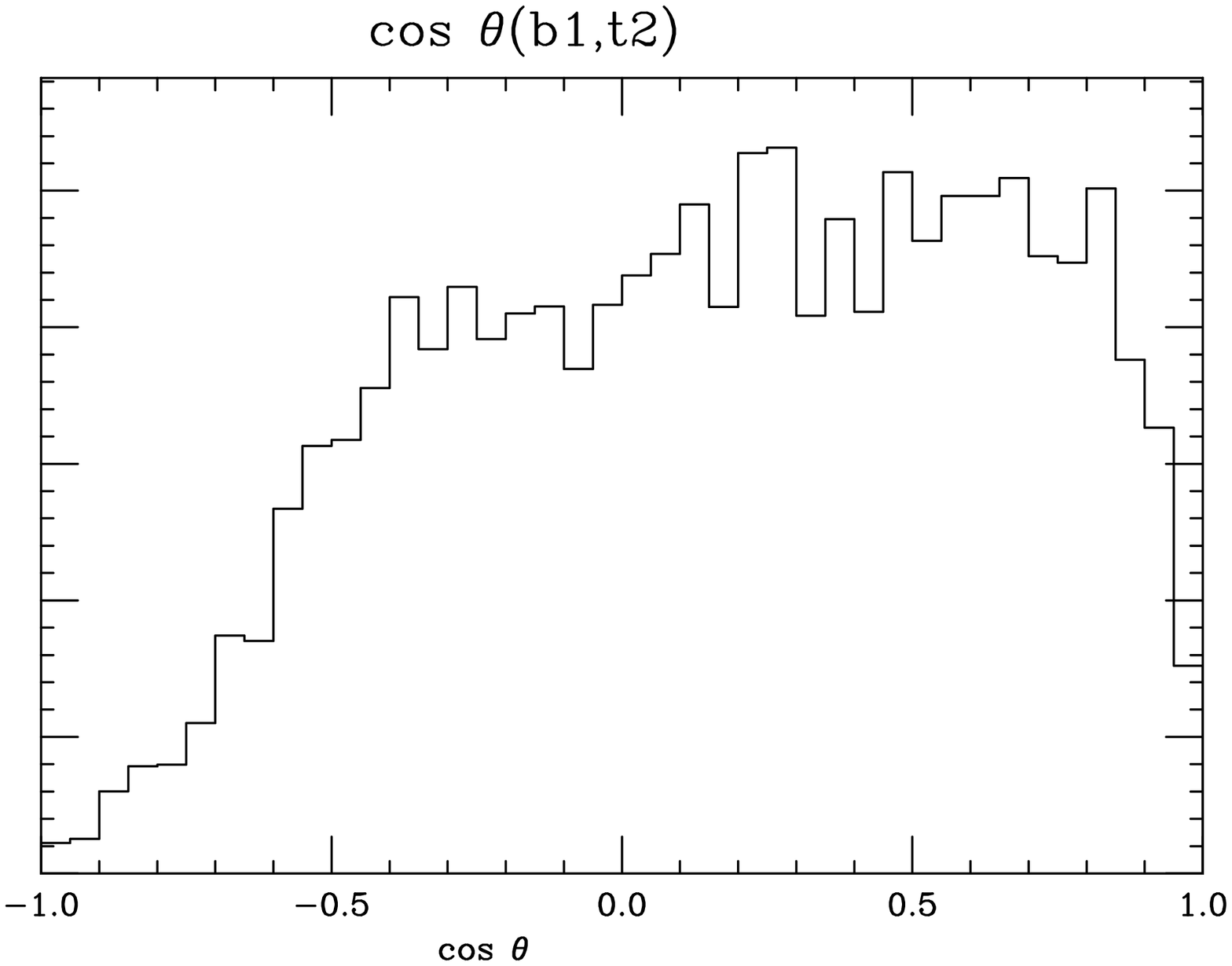}
\includegraphics[angle=90,width=3.0in]{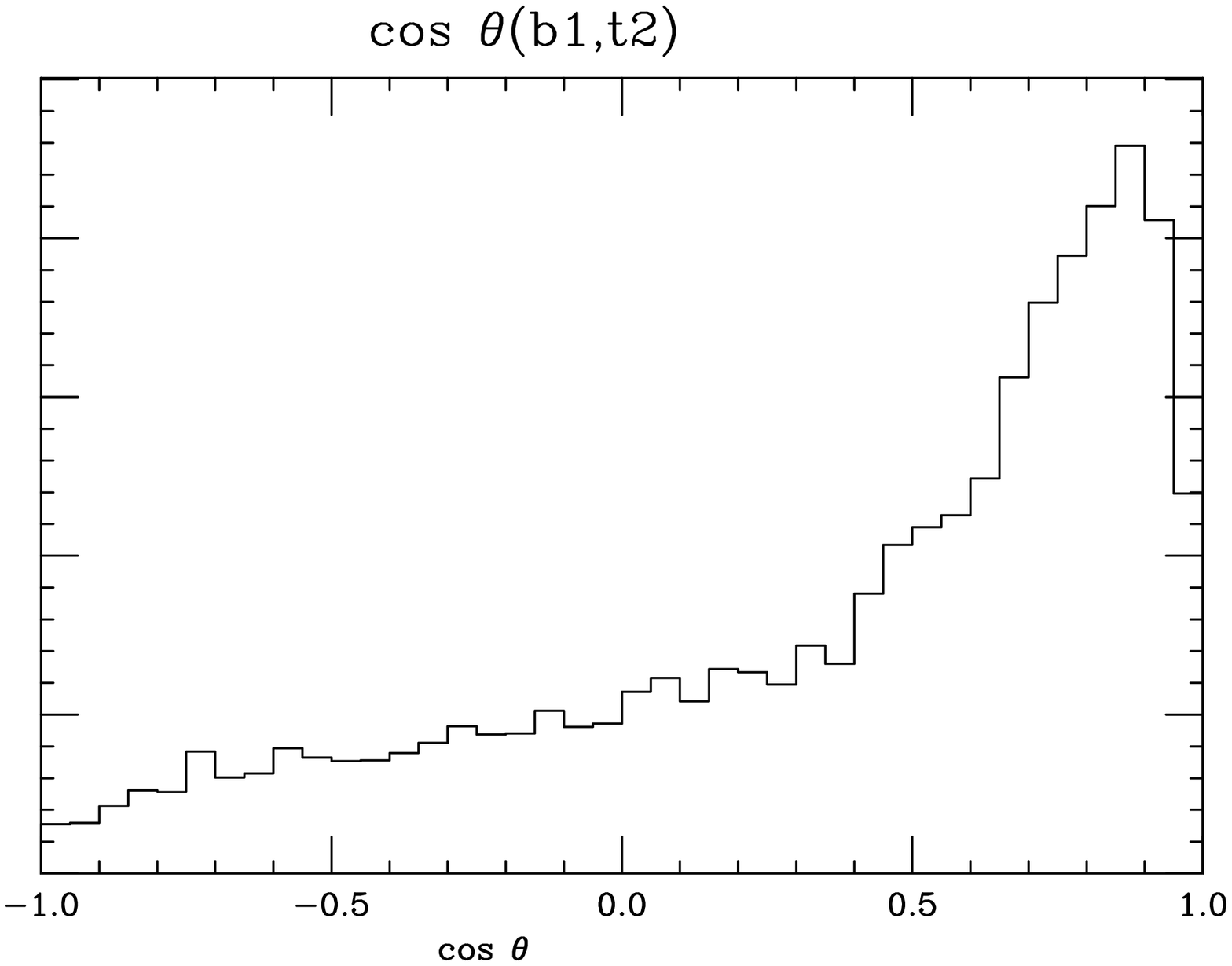}
\caption{ Lab frame polar angle ${\rm cos}\, \theta_{t_2 b_1}$ for the charged octet signal (left panel) and the $t\bar{t}j$ background (right panel).  Included are the two top tags, one $b$-tag, and
$H_T>1000\;{\rm GeV}$ cut of Table~\ref{kincuts}.  The normalization of the vertical axis is arbitrary.  All scales on each plot are linear.}
 \label{ctplot}
\end{figure}

We can improve on the significance of the signal by using details of the event shape.  The signal consists of a prompt top quark recoiling against the color octet scalar, which decays to another top quark and a $b$-jet.  Since the scalar is heavy and is not produced with a large boost, its decay products  will not
typically be highly collimated with the prompt top quark.  We expect the two top quarks and the $b$-jet to form a ``Mercedes Benz" topology, as shown in Fig.~\ref{MB}.  However, we find that the $t\bar{t}j$ events that survive the above kinematic cuts have a very back-to-back $t_1$, $b_1$ pair.  This motivates the further set of cuts in
the lower half of Table~\ref{kincuts}.  We have used $\theta_{t_i b_1}$ to denote the lab frame angle between the top jets and the $b$-jet.  These cuts are designed to select events with the signal
characteristics described above.  For illustration of the kinematic differences between signal and background, the ${\rm cos}\, \theta_{t_2 b_1}$ distribution for both with
arbitrary normalization is shown in Fig.~\ref{ctplot}.

After imposing these cuts, we find the values of $S/\sqrt{B}$ and $S/B$ listed in Table~\ref{sigs}.  The successive improvement in both measures is evident as the cuts are imposed.  We note that the large significances obtained with only the $H_T$ cut are large because only the $t\bar{t}j$ background has been considered.  We believe that demanding the top and $b$ tags is crucial in reducing other backgrounds in the fully hadronic mode.
It may be possible to relax the tagging requirements in the semi-leptonic mode.
Observation of this signal with $5\sigma$ significance requires slightly more than $300 \,{\rm fb}^{-1}$ if only semi-leptonic events are used, while less than $100 \,{\rm fb}^{-1}$ is sufficient for discovery if hadronic tops can be implemented in the analysis.
We note that
this is for the value $\eta_U=1$; the significance $S/\sqrt{B}$ scales as $|\eta_U|^2$, while the required luminosity for $5\sigma$ observation scales as $1/|\eta_U|^4$.
Unless $\eta_U>1$, this is a search mode that requires a large amount of integrated luminosity.  The purity of the sample after cuts, as measured by $S/B$, is reasonably high.  We have not used all possible kinematic handles, so these results
could likely be improved somewhat.  For example, the invariant mass and $p_T$ distributions of the top quarks and bottom jet could be binned and compared between
signal and background, allowing the shape differences to be more fully exploited.  Our goal here is only to show that observation of the signal is feasible at the LHC.

\begin{table}[htbp]
\centering
\begin{tabular}{c c c c c}
\hline\hline
Cut & ${S_l}/{\sqrt{B_l}}$ & ${S_l}/{B_l}$ & ${S_{l+h}}/{\sqrt{B_{l+h}}}$ & ${S_{l+h}}/{B_{l+h}}$ \\ [1ex]
\hline
$H_T > 1000\:{\rm GeV}$ & 5.52 & 0.0062 & 8.89  & 0.0062  \\ [1ex]
$\exists \,t_1,t_2,b_1$ & 0.95 & 0.0096 & 1.92 & 0.0099  \\[1ex]
$900\:{\rm GeV}<M_{b_1,t_1}<1100\:{\rm GeV}$ & 2.60  & 0.12 & 5.93 & 0.16 \\ [1ex]
\hline
$\Delta R_{b_1 t_1}, \Delta R_{b_1 t_2}< 3.0$ & 3.60 & 0.32 & 8.82 & 0.47 \\ [1ex]
${\rm cos}\, \theta_{t_1 b_1}<0.5,{\rm cos}\, \theta_{t_2 b_1}<0.7$  & 4.56 & 0.62 & 10.3 & 0.73 \\ [1ex]
\hline
\end{tabular}
\caption{\label{sigs} Obtained significances after successively imposing the cuts in Table~\ref{kincuts}.  $S_{l+h}$ denotes the combination of semi-leptonic and fully hadronic signal events, and
$B_{l+h}$ denotes the same for the background.  We assume $300\,{\rm fb}^{-1}$ of integrated luminosity; other values of $L$ change only $S/\sqrt{B}$, and can be obtained from the numbers
in the table by the scaling $\sqrt{L/300\;{\rm fb}^{-1}}$. }
\end{table}

An observation of signal over background will permit measurement of the coupling $|\eta_U|^2$.
$\eta_U$ can be expressed via GUT-scale parameters using Eqs.~(\ref{etadef}) and~(\ref{guteta}).
The obtainable precision on this parameter depends 
on statistical uncertainties, and on how well the signal and background cross sections can be predicted.  We assume that these errors can be approximately written as fractional uncertainties in the signal and background cross
sections, in which case the coupling uncertainty is
\begin{equation}
\frac{\delta |\eta_U|^2}{|\eta_U|^2} = \frac{\sqrt{S+B}\:\oplus F_S S \:\oplus F_B B}{S}.
\label{sigun}
\end{equation}
$F_S$ and $F_B$ denote the fractional uncertainties for the $S^{\pm}t$ and $t\bar tj$ cross sections, respectively.  We add the various components of the
error in quadrature.  We consider uncertainties arising from higher-order QCD effects, and begin our estimate of the errors by varying the scale choice in our calculation by a factor of two around our canonical
value $\mu=\sqrt{m_t^2+\sum p_{T}^2}$ for both signal and background.
These variations give some measure of the uncertainties in the leading order cross sections, which we denote as $F^\text{LO}_{S,B}$.
For the inclusive cross section, we find $F^\text{LO}_S\approx 0.15$ and $F^\text{LO}_B \approx 0.40$.
The size of the variations are similar after the cuts are imposed, although the statistics and severity of the cuts for the background make it difficult to get a very precise estimate.
Inserting these numbers into Eq.~(\ref{sigun}), and assuming $300\,{\rm fb}^{-1}$, leads to $\delta S/S = 0.72$ for only semi-leptonic decays and
$\delta S/S = 0.58$ when hadronic decays are added.  Not only is a precise determination of the color-octet Yukawa impossible,
it would be difficult to ascertain that a signal was even present.

The above discussion clearly indicates that knowledge of the higher-order QCD corrections to both signal and background is required for interpretation of the measurement.
The next-to-leading order QCD corrections to the signal are unknown, but are straightforward to calculate.  We study below the
precision in the signal rate that can be obtained assuming $F_S=0.05$ and $F_S=0.10$, which we consider reasonable estimates of the residual error after
next-to-leading order QCD corrections are included.  The next-to-leading order QCD corrections to the $t\bar{t}j$ background are
known~\cite{Dittmaier:2007wz}.  However, a flexible numerical program that allows the imposition of phase-space constraints has not yet been released.  In lieu of a
better solution, we estimate the uncertainty of our background using the inclusive cross section.
We first note using Ref.~\cite{Dittmaier:2007wz} that varying the leading-order cross section about a fixed scale by a factor of 2 also gives an approximate uncertainty of $\pm40\%$, as does the variation about the running scale used in our analysis.
We then estimate that the residual scale variation of the next-to-leading order result is $\pm 10$--$20\%$.
Other handles
on the normalization of this background will
be available at the LHC.  Significant statistics outside the signal region exist to normalize simulation predictions and determine the partonic luminosities that
contribute.  We consider below $F_B=0.10$ and $F_B=0.20$ as estimates of the obtainable precision in the background prediction.  We have not included experimental systematics 
in our error estimate; these will certainly affect our results.  It should be possible to gain control over these effects using the significant statistics outside the signal region.

We present in Table~\ref{smeas} the expected accuracy in the measurement of the coupling for the various error estimates
discussed above.  The limiting
statistical precision for $\eta_U=1$ assuming $300\,{\rm fb}^{-1}$ is $\pm 14\%$ if all top modes can be used, and $\pm 28\%$ if only the semi-leptonic decays are included.
With the possibly optimistic assumption that the background can be controlled with $\pm10\%$ precision via a combination of next-to-leading order QCD and extrapolation from signal-free phase space regions, a precision of 19\% in the color-octet coupling is possible with $300\,{\rm fb}^{-1}$ and $\eta_U=1$ using all top modes.
This becomes 8\% if $\eta_U=2$.  The more pessimistic assumption of $\pm 20\%$ precision leads
to measurements of 31\% precision for $\eta_U=1$ and 10\% for $\eta_U=2$.  We caution the reader again regarding several omissions in our
limited analysis; inclusion of experimental systematics will degrade the obtainable precisions, while more thoroughly using the shape differences between signal and background will improve
the results.

\begin{table}[htbp]
\centering
\begin{tabular}{cc|cc|cc}
\hline\hline
$F_S$ & $F_B$ & $\frac{\delta |\eta_U|^2}{|\eta_U|^2}$ (semi)  & $\frac{\delta |\eta_U|^2}{|\eta_U|^2}$ (all) & $\frac{\delta |\eta_U|^2}{|\eta_U|^2}$ (semi)  & $\frac{\delta |\eta_U|^2}{|\eta_U|^2}$ (all)\\ [1ex]
\hline
\multicolumn{2}{c|}{} & \multicolumn{2}{c|}{$\eta_U=1$} &  \multicolumn{2}{c}{$\eta_U=2$} \\ [1ex]
\hline
0   &    0 &  0.28 & 0.14 & 0.10 & 0.05 \\[1ex]
0.05 & 0.10 & 0.33 & 0.19 & 0.12 & 0.08 \\[1ex]
0.05 & 0.20 & 0.43 & 0.31 & 0.14 & 0.10 \\[1ex]
0.10 & 0.10 & 0.34 & 0.21 & 0.15 & 0.12\\[1ex]
0.10 & 0.20 & 0.44 & 0.32 & 0.16 & 0.13 \\[1ex]
0.15 & 0.40 & 0.72 & 0.58 & 0.24 & 0.21 \\[1ex]
\hline
\end{tabular}
\caption{\label{smeas}Accuracy of signal measurement for single charged scalar production assuming various estimates of the errors in signal and background predictions.  We have included
results assuming $300\,{\rm fb}^{-1}$ of integrated luminosity for the choices $\eta_U=1,2$.  Included are results using only semi-leptonic events (labeled semi) and
results obtained using all events (labeled all).}
\end{table}

\subsection{$b\bar{b} \to S_{R,I} \to t\bar{t}$ resonant production}
Another possible mode providing insight to GUT physics is resonant production of the neutral scalars through the partonic process $b\bar{b} \to S_{R,I} \to t\bar{t}$.
This process is proportional to the coupling combination $\eta_{bb} \equiv |\eta_U \eta_D|^2/\Gamma_S$.
This production mode is normally suppressed by $m_b^2/\upsilon^2$, but can become sizable with large couplings.  We take
$\eta_D=40$ and $\eta_U=1$ to simulate this possibility.  We note that the widths depend differently on $\eta_{U,D}$ than the cross section, so different choices that leave
the product $|\eta_U \eta_D|$ unchanged will lead to different results.

An important constraint on the parameter $|\eta_U \eta_D|$ is the decay $b \to s\gamma$.  Within the framework of minimal flavor violation, the overall coupling for this decay goes like
$\eta_U \eta_D V_{tb}$, where $V_{tb}$ denotes the $(3,3)$ element  of the CKM matrix.  For $\eta_D=40$, the constraints arising from $b\to s\gamma$ on $m_S$ within minimal flavor violation are several TeV~\cite{Manohar:2006ga}, pushing this state beyond the reach of the LHC.  If minimal flavor violation is not assumed, then $V_{tb}$ is replaced with an element of an arbitrary mixing matrix, which can be much smaller than $V_{tb} \approx 1$, permitting $m_S = 1\,{\rm TeV}$.  We assume $m_S=1\,{\rm TeV}$ to allow for this possibility at the LHC.

Our study of this channel proceeds similarly to the charged scalar analysis.
The leading order cross section for this process is 290 fb, several times larger than the loop-induced process $gg \to S_{R,I} \to t\bar{t}$ studied in Ref.~\cite{Gresham:2007ri}.
We focus on the $t\bar{t}$ background, again relying on previous studies that have indicated that other backgrounds can be suppressed in the semi-leptonic mode~\cite{tdrs,Barger:2006hm} and with the previously discussed caveats on the fully hadronic mode.
The cross section at leading order for $t\bar{t}$ production at the LHC is 640 pb.
We present results for both the semi-leptonic and fully hadronic channels, as previous studies have shown that suppression of QCD multi-jet backgrounds is possible~\cite{Thaler:2008ju,Kaplan:2008ie}.
We slightly tweak the parameters in our top reconstruction algorithm to account for the different kinematics of this process with respect to the single charged scalar production: we set $\Delta R_\text{max}=0.8$ in the semi-leptonic top reconstruction and $\Delta R_\text{max}=1.1$ in the hadronic top reconstruction.
We study this mode using a sample of $10^4$ signal events and $10^6$ $t\bar{t}$ background events.

\begin{table}[htbp]
\centering
\begin{tabular}{c c c c c}
\hline\hline
Cut & $S_l$ & $S_h$ & $B_l$ & $B_h$\\ [1ex]
\hline
$H_T > 900\:{\rm GeV}$ & 1500 & 4930 & $6.22\times10^4$ & $1.96\times10^5$ \\ [1ex]
$\exists \,t_1,t_2$ & 494 & 2840 & $2.37\times10^4$ & $1.20\times10^5$ \\[1ex]
$950\:{\rm GeV}<M_{t_1,t_2}<1050\:{\rm GeV}$ & 374 & 2380 & 5050 & $2.81\times10^4$ \\ [1ex]
\hline
\end{tabular}
\caption{\label{kincuts2} Effects of kinematic cuts on cross sections.  $S_l$ and $S_h$ denote the expected numbers of semi-leptonic and hadronic signal events after cuts for $100 \,{\rm fb}^{-1}$ of integrated luminosity; $B_l$ and $B_h$ denote the same for the background.}
\end{table}

We present in Table~\ref{kincuts2} the effect of the reconstruction on the signal and background processes assuming $100\,{\rm fb}^{-1}$.  We demand $H_T>900\,{\rm GeV}$ and require that our top reconstruction algorithm find two top candidates.
We also demand that the invariant mass reconstruct to the scalar mass.
The results for $S/\sqrt{B}$ and $S/B$ obtained after each successive cut are shown in Table~\ref{sigs2}; we again note that the large significances obtained with only the $H_T$ cut are illusory, as we believe the top tags are required to suppress other sources of background, at least in the fully hadronic mode.
Our results indicate that observation is possible using only the semi-leptonic mode with $100\,{\rm fb}^{-1}$; the significance is over 10 if the hadronic events can be used.
We present in Table~\ref{smeas2} a study of the estimated accuracy of the signal measurement using Eq.~(\ref{sigun}).
It is clear from the $S/B$ values appearing in Table~\ref{sigs2} that control of the systematic error in the background will be crucial.
As it should be possible to extrapolate the background from the $M_{t_1,t_2}$ sidebands, we study the expected accuracy assuming small normalization uncertainties in the background ranging from 1--5\%.
If a 1\% normalization uncertainty of the background is possible, then a measurement of the $\eta_{bb}$ coupling of 24\% accuracy using only semi-leptonic modes is possible.  This improves
to 14\% if all top modes can be utilized.  The estimated precision worsens dramatically as the uncertainty $F_B$ is increased.

\begin{table}[htbp]
\centering
\begin{tabular}{c c c c c}
\hline\hline
Cut & ${S_l}/{\sqrt{B_l}}$ & ${S_l}/{B_l}$ & ${S_{l+h}}/{\sqrt{B_{l+h}}}$ & ${S_{l+h}}/{B_{l+h}}$ \\ [1ex]
\hline
$H_T > 900\:{\rm GeV}$ & 6.0 & 0.024 & 12.7  & 0.025  \\ [1ex]
$\exists \,t_1,t_2$ & 3.2 & 0.021 & 8.8 & 0.023  \\[1ex]
$950\:{\rm GeV}<M_{t_1,t_2}<1050\:{\rm GeV}$ & 5.3  & 0.074 & 15.1 & 0.083 \\ [1ex]
\hline
\end{tabular}
\caption{\label{sigs2} Obtained significances for the neutral scalar resonance production after successively imposing the cuts in Table~\ref{kincuts2}.  $S_{l+h}$ denotes the combination of semi-leptonic and fully hadronic signal events, and
$B_{l+h}$ denotes the same for the background.  We assume $100\,{\rm fb}^{-1}$ of integrated luminosity.}
\end{table}

\begin{table}[htbp]
\centering
\begin{tabular}{c|cc}
\hline\hline
$F_B$ & $\frac{\delta \eta_{bb}}{\eta_{bb}}$ (semi)  & $\frac{\delta \eta_{bb}}{\eta_{bb}}$ (all)\\ [1ex]
\hline
0 & 0.20 & 0.07  \\[1ex]
0.01 & 0.24 & 0.14  \\[1ex]
0.025 & 0.39 & 0.31  \\[1ex]
0.05 & 0.70 & 0.60  \\[1ex]
\hline
\end{tabular}
\caption{\label{smeas2} Accuracy of signal measurement for the neutral scalar resonances assuming various estimates of the theoretical errors in signal and background predictions.  We have included
results assuming $100\,{\rm fb}^{-1}$ of integrated luminosity for the choice $\eta_U=1$, $\eta_D=40$.  Included are results using only semi-leptonic events (labeled semi) and
results obtained using all events (labeled all).}
\end{table}

\section{Conclusions \label{sec:conc}}
In this paper we have studied the production and detection of
color-octet scalars at the LHC.  We have focused on single production of both
charged and neutral members of an $({\bf 8,2})_{1/2}$ doublet through
bottom-quark initiated partonic processes. Such production modes probe
the Yukawa structure of the scalar sector. These scalars appear in realistic $SU(5)$
grand unified theories where the ${\bf 45}_H$ representation is used to generate fermion
masses for the charged fermions.  One of their couplings is defined by the 
difference between the down quark and charged lepton Yukawa couplings.  In the case of adjoint $SU(5)$~\cite{adjoint}, these fields are 
expected to be light to satisfy constraints coming from unification and proton decay, and may have TeV-scale masses.  Measuring the mass of the color-octet scalar 
gives an upper bound on the proton lifetime.  Thus, the LHC may potentially be able to probe GUT-scale physics if these states are observed.  

We study observation of TeV-mass
scalars using the two partonic modes $bg \to S^{\pm} t \to t\bar{t} b$ and
$b\bar{b} \to S_{R,I} \to t\bar{t}$.
We carefully analyze whether the signal can be seen over Standard Model top-quark backgrounds at the LHC, and devise a set of cuts based on the kinematics of the signal that facilitate observation.
As the final state contains boosted top quarks, we discuss their reconstruction in both semi-leptonic and hadronic decay modes.
We find that observation of the charged scalar with $S/\sqrt{B} \geq 5$ is possible with slightly more than $300\;{\rm fb}^{-1}$ if only semi-leptonic $t\bar{t}$ modes are used, while less than $100\;{\rm fb}^{-1}$ is needed if fully hadronic top decays can be utilized.
Observation of the  $b\bar{b} \to S_{R,I} \to t\bar{t}$
channel requires less than $100\;{\rm fb}^{-1}$ in both cases.
We roughly estimate the effect of systematic errors on the measurement of the Yukawa parameters that lead to the signal
cross section, and find obtainable precisions of 10--30\%, depending on the size of the couplings.

In summary, probing the color-octet scalar Yukawa structure through bottom-quark initiated processes at the LHC may 
potentially provide a window to GUT-scale physics, and create an interesting synergy between 
collider measurements and future proton-decay searches.

\begin{acknowledgments}
RG, TM and FP are supported by the DOE grant DE-FG02-95ER40896,
Outstanding  Junior Investigator Award, by the Wisconsin
Alumni Research Foundation, and by the Alfred P.~Sloan Foundation.
The work of PFP was supported in part by the U.S.~Department of
Energy contract No. DE-FG02-08ER41531 and in part by the Wisconsin Alumni
Research Foundation.
\end{acknowledgments}

\appendix
\section{Decay and Production Formulae}
We present in this Appendix the production and decay formulae for the charged and neutral scalars used in our analysis.  We begin with the decay modes.  The decay rates for $S_R$ are
\begin{eqnarray}
\Gamma (S_R \to b\bar{b}) &=& \frac{m_S}{16 \pi}\frac{|\eta_D|^2 m_b^2}{v^2},\nonumber\\
\Gamma (S_R \to t\bar{t}) &=& \frac{m_S}{16 \pi}\frac{|\eta_U|^2 m_t^2}{v^2}\Bigg\{1-\frac{4m_t^2}{m_S^2}\Bigg\}^{3/2},\nonumber\\
\Gamma_R &=& \frac{m_S}{16 \pi}\Bigg\{\frac{|\eta_D|^2 m_b^2}{v^2}+\frac{|\eta_U|^2 m_t^2}{v^2}\Bigg{(}1-\frac{4m_t^2}{m_S^2}\Bigg{)}^{3/2}\Bigg\}.
\end{eqnarray}
The decay rates for $S_I$ are
\begin{eqnarray}
\Gamma (S_I \to b\bar{b}) &=& \frac{m_S}{16 \pi}\frac{|\eta_D|^2 m_b^2}{v^2},\nonumber\\
\Gamma (S_I \to t\bar{t}) &=& \frac{m_S}{16 \pi}\frac{|\eta_U|^2 m_t^2}{v^2}\Bigg\{1-\frac{4m_t^2}{m_S^2}\Bigg\}^{1/2},\nonumber\\
\Gamma_I &=& \frac{m_S}{16 \pi}\Bigg\{\frac{|\eta_D|^2 m_b^2}{v^2}+\frac{|\eta_U|^2 m_t^2}{v^2}\Bigg{(}1-\frac{4m_t^2}{m_S^2}\Bigg{)}^{1/2}\Bigg\}.
\end{eqnarray}
The decay rate for the charged scalar is
\begin{equation}
\Gamma(S^+ \to t\bar{b}) = \frac{1}{16 \pi m_S^3}  \frac{|\eta_U|^2 m_t^2}{v^2} (m_S^2-m_t^2)^2.
\end{equation}
As we only study the charged scalar in the limit $\eta_D =1$, we neglect contributions from this coupling.

We now present the partonic cross sections for the relevant processes.
The differential partonic cross section for charged-scalar pair production via gluon fusion studied in Fig.~\ref{csecs} is
\begin{equation}
\frac{d\hat\sigma(gg\to S^+S^-)}{dt}=\begin{aligned}[t]
&\frac{9\pi\alpha_s^2}{4\hat s^4}\left(\frac{\hat s^2-m_S^2\hat s+m_S^4+\hat st-2m_S^2t+t^2}{(\hat s+t-m_S^2)^2(t-m_S^2)^2}\right)\\
&\times\big(\begin{aligned}[t]
&\hat s^2t^2+m_S^4\hat s^2+2m_S^4\hat st+2\hat st^3-4m_S^2\hat st^2+t^4\\
&+6m_S^4t^2-4m_S^6t+m_S^8-4m_S^2t^3\big),
\end{aligned}
\end{aligned}
\end{equation}
where the Mandelstam variable $t=(p_{g_1}-p_{S^+})^2=(p_{g_2}-p_{S^-})^2$, and $\hat{s}$ is the partonic center of mass energy.
The differential partonic cross section of $S^+\bar{t}$ production from $g\bar{b}$ scattering is
\begin{equation}
\frac{d\hat\sigma(g\bar b\to S^+\bar t)}{dt}=\begin{aligned}[t]
&\frac{\alpha_s}{72\hat s^3}\frac{|\eta_U|^2m_t^2}{v^2}\left(\frac{4t^2-\hat st+4\hat s^2-8tm_t^2+\hat sm_t^2+4m_t^4}{(t-m_t^2)^2(\hat s+t-m_t^2)^2}\right)\\
&\times\big(\begin{aligned}[t]
&m_t^2\hat s^2-\hat s^2t+2m_t^2\hat st-2\hat st^2-2m_t^4\hat s+2m_S^2\hat st-2m_S^4t+m_t^6-t^3\\
&+2m_S^2t^2+2m_S^4m_t^2-2m_S^2m_t^4+m_t^2t^2-m_t^4t\big),
\end{aligned}
\end{aligned}
\end{equation}
where now the Mandelstam variable $t=(p_g-p_{\bar{t}})^2=(p_{\bar{b}}-p_{S^+})^2$.
The differential partonic cross section for resonant top-quark pair production mediated by the neutral octet scalar, $b\bar{b} \to S_{R,I} \to t\bar{t}$, is as follows.  The interference between the
$S_R$ and $S_I$ exchanges vanishes, and the cross section conveniently separates into two simple expressions:
\begin{eqnarray}
\frac{d\hat{\sigma}(b\bar{b} \to S_R \to t\bar{t})}{dt} &=& \frac{1}{72\pi} |\eta_U \eta_D|^2 \frac{m_t^2 m_b^2}{v^4} \frac{1}{(\hat{s}-m_S^2)^2+m_S^2 \Gamma_R^2} \left(1-\frac{4 m_t^2}{\hat{s}}\right),\\
\frac{d\hat{\sigma}(b\bar{b} \to S_I \to t\bar{t})}{dt} &=& \frac{1}{72\pi} |\eta_U \eta_D|^2 \frac{m_t^2 m_b^2}{v^4} \frac{1}{(\hat{s}-m_S^2)^2+m_S^2 \Gamma_I^2},
\end{eqnarray}
where $\Gamma_R$ and $\Gamma_I$ have been defined above.

\end{document}